\documentclass[aps,pre,superscriptaddress,twocolumn,showpacs,showkeys,longbibliography]{revtex4}

\usepackage{graphicx,epstopdf,url}
\usepackage[colorlinks=true,urlcolor=blue,citecolor=blue,linkcolor=blue,urlcolor=blue]{hyperref}

\usepackage[active]{srcltx}

\begin{document}
\title{Hydroxygraphene: dynamics of hydrogen bond networks}

\author{Alexander V. Savin}
\affiliation{ N.N. Semenov Federal Research Center for Chemical Physics,
Russian Academy of Science (FRCCP RAS), Moscow, 119991, Russia}
\affiliation{Plekhanov Russian University of Economics, Moscow, 117997 Russia}

\begin{abstract}

Using the molecular dynamics method, dynamics of hydrogen bond (HB) networks emerging on the surface
of a graphene sheet during its functionalization with hydroxyl groups OH are simulated.
It is demonstrated that two OH groups form an energetically more advantageous structure
when they are covalently attached on one side of the sheet to carbon atoms forming opposite vertices
of one hexagon of valence bonds of the sheet.
Attaching of OH groups to carbon atoms located at the opposite vertices of hexagons
of valence bonds leads to the emergence of hydroxygraphene C$_4$(OH).
In such sheet lying on a flat substrate, attached  oxygen atoms on its outer surface form a
hexagonal lattice, and hydroxyl groups due to their turns can in various ways form chains of hydrogen
bonds. The modification of the sheet from two sides results in forming of hydroxygraphene C$_2$(OH)
with HB networks on both sides of the graphene sheet.
Simulation of the dynamics of these sheets shows that their heat capacity at low temperatures
$T<T_0$ increases monotonously when the temperature rises, reaches its maximum at $T=T_0$
and then decreases monotonically.
The initial growth is caused by the accumulation of orientational defects in the lattice
of hydrogen bonds whereas the decrease at $T>T_0$ is explained by the "melting"\ of the lattice.
For one chain of OH groups connected to the outer side of a nanoribbon the melting temperature is
$T_0=500$K, while for a graphene sheet C$_4$(OH) modified on one side $T_0=260$K,
and for a graphene sheet C$_2$(OH) modified on both sides $T_0=485$K.

\end{abstract}

\maketitle

\section{Introduction}

Molecular systems with chains of hydrogen bonds (HB) have been the object of many scientific
investigations in physics, chemistry and biology in the last decades.
Important goal of studying such systems is the proton transport that takes place
through the hydrogen bonds \cite{nt1983mb,k1996cm,nvvtat2011cc}.
The proton conductivity in HB networks is very high \cite{z2000acm}.
The chains of hydrogen bonds act as proton wires providing an effective pathway for the
rapid transportation of protons.  Quantum mechanical aspects of proton transfers in
quasi-one dimensional HB systems have been studied in \cite{f2002jms},
nonlinear (topological solitary waves) aspects --- in \cite{kycy2004pre}.
Experimental and theoretical studies of water-filled carbon nanotubes have shown
that water molecules can penetrate opened nanotubes and form inside extended chains
of hydrogen bonds \cite{hrn2001n,dnh2003prl}.
The mechanisms of proton transport in these chains was discussed in \cite{kycy2004pre,clzmw2013pccp}.

Investigations \cite{mrrs2010acsn,lzltz2018lmm} show that the large-scale properties
of graphene oxide (GO) platelets are controlled by hydrogen bond networks that involve functional
groups on GO layers (such as epoxy\linebreak ---O--- and hydroxy ---OH groups)
and water molecules OH$_2$ within the interlayer cavities.
An increase in the number of water molecules between layers of GO leads to an increase
in its proton conductivity \cite{rky2018pra}.
Graphane functionalized with hydroxyl groups OH (hydroxygraphane C$_2$H(OH)) \cite{wk2010njp}
can conduct protons in the complete absence of water through a contiguous network of hydrogen bonds
\begin{equation}
\begin{tabular}{lllll}
OH$\cdots$&OH$\cdots$&OH$\cdots$&OH$\cdots$&OH$\cdots$ \\
~$|$ & ~$|$ & ~$|$ & ~$|$ & ~$|$ \\
C & C & C & C & C
\end{tabular}
\label{f1}
\end{equation}
Density functional theory calculations predict remarkably low barriers to diffusion of proton
along 1D chain of surface hydroxyl groups \cite{bcsdgj2017prl}.
The properties of hydroxygraphane due to the presence of surface network of hydrogen bonds
\cite{bj2019jpcl} make it a viable candidate for a proton exchange membrane material capable
of operating under anhydrous or low-humidity conditions.

In this work, we investigate the structures and the dynamics of HB networks emerging
on the surface of graphene sheet by its functionalization with hydroxyl groups OH.
It will be shown that two OH groups form an energetically more advantageous configuration
when they are attached on the one side of the sheet to the carbon atoms at the opposite vertices
of one hexagon of valence bonds of graphene (in this case the attached groups can form
a full-fledged hydrogen bond OH$\cdots$OH).
After such one side functionalization of the graphene sheet we will have hydroxygraphene
with the formula C$_4$(OH) (one attached group per four carbon atoms).
In this structure, atoms of oxygen O on the surface of the sheet will be located
in the points of the hexagonal lattice, whereas OH groups can in various ways form zigzag chains
of hydrogen bonds (\ref{f1}). A two-sided modification of a sheet results in obtaining
hydroxygraphene C$_2$(OH) with HB networks on both sides of the graphene sheet.

\section{Model of modified graphene sheet}
\label{Sec2}

To model the chemically modified graphene sheet, we use the force field in which
distinct potentials describe the deformation of valence bonds and valence, torsion and dihedral
angles, and non-valence atomic interactions \cite{swcb1995acc,skh2010prb}.
In this model, the deformation energy of the valence sp$^2$ and sp$^3$ C--C and C--COH bonds,
and of C--H, C--OH, O--H bonds is described by the harmonic potential:
\begin{equation}
V(\rho)=\frac12 K(\rho-\rho_0)^2,
\label{f2}
\end{equation}
where $\rho$ and $\rho_0$ are the current and equilibrium bond lengths,
$K$ is the  bond stiffness. The values of potential parameters for various
valence bonds are presented in Table~\ref{tab1}.

Energies of the deformation of the valence angles X--Y--Z are described by the potential
\begin{equation}
U({\bf u}_1,{\bf u}_2,{\bf u}_3)=U(\varphi)=\epsilon_{a}(\cos\varphi-\cos\varphi_0),
\label{f3}
\end{equation}
where the cosine of the valence angle is defined as
$\cos\varphi=-({\bf v}_1, {\bf v}_2)/|{\bf v}_1 || {\bf v}_2 |$,
with the vectors ${\bf v}_1 = {\bf u}_2 - {\bf u}_1 $, ${\bf v}_2 = {\bf u}_3 - {\bf u}_2$,
the vectors ${\bf u}_1$, ${\bf u}_2 $, ${\bf u}_3 $ specify the coordinates of the atoms forming
the valence angle $\varphi$, $\varphi_0$ is the value of equilibrium valence angle.
The values of potential parameters used for various valence angles are presented in Table~\ref{tab2}.

Deformations of torsion and dihedral angles, in the formation of which edges carbon atoms
with attached external atoms do not participate (torsion angles around the sp$^2$ C--C bonds),
are described by the potential:
\begin{equation}
W_1({\bf u}_1,{\bf u}_2,{\bf u}_3,{\bf u}_4)=\epsilon_{t,1}(1-z\cos\phi),
\label{f4}
\end{equation}
where $\cos\phi=({\bf v}_1,{\bf v}_2)/|{\bf v}_1||{\bf v}_2|$, with the vectors
${\bf v}_1=({\bf u}_2-{\bf u}_1)\times ({\bf u}_3-{\bf u}_2)$,
${\bf v}_2=({\bf u}_3-{\bf u}_2)\times ({\bf u}_3-{\bf u}_4)$,
the  factor $z=1$ for the dihedral angle (the equilibrium angle $\phi_0=0$)
and  $z=-1$ for the torsion angle (the equilibrium angle $\phi_0=\pi$),
$\epsilon_{t,1}=0.499$ eV is the binding energy (the vectors ${\bf u}_1$,...,${\bf u}_4$
specify the coordinates of the atoms, which form the angle).
More detailed description of the energy of deformation of the torsion and dihedral angles
is given in \cite{skh2010prb}.
\begin{table}[t]
\caption{Values of the harmonic potential parameters (\ref{f2}) for different
valence bonds X---Y (C and C$'$ are carbon atoms involved in the formation of the
sp$^2$ and sp$^3$ bonds). \label{tab1}
}
\begin{center}
\begin{tabular}{l|ccccc}
 ~X---Y~    & ~~C--C~~ &  C--C$'$, C$'$--C$'$ &  ~~C--H~~ & ~~C--O~~ &  ~~O--H~~\\
\hline
~$K$~(N/m)~ & 508.9 & 348.9 & 444.3 & 349.8 & 444.3 \\
~$\rho_0$~(\AA)& 1.418 & 1.522 & 1.08  & 1.41  & 0.96\\
\hline
\end{tabular}
\end{center}
\end{table}

Deformations of the angles around sp$^3$ bonds C--C$'$, C$'$--C$'$ are described by
the potential:
\begin{equation}
W_2({\bf u}_1,{\bf u}_2,{\bf u}_3,{\bf u}_4)=\epsilon_{t,2}(1+\cos3\phi),
\label{f5}
\end{equation}
with binding energy $\epsilon_{t,2}=0.03$ eV.

The nonvalent van der Waals interactions of atoms are described by the Lennard-Jones potential
\begin{equation}
W_0(r)=\epsilon_0[(r_0/r)^{12}-2(r_0/r)^6],
\label{f6}
\end{equation}
where $r$ and $r_0$ are the current and equilibrium distance between interacting atoms,
$\epsilon_0$ is the interaction energy. For atoms C and O we use the values
$\epsilon_0=0.00658$~eV, $r_0=3.629$~\AA, for atoms C and H --
$\epsilon_0=0.0018$~eV, $r_0=3.395$~\AA.
\begin{table}[t]
\caption{Values of the parameters of the potential of the valence angle X--Y--Z (\ref{f3}) for different
atoms. \label{tab2}
}
\begin{center}
\begin{tabular}{l|ccccc}
 ~~X--Y--Z~~              & ~C--C--C~ & ~C--C$'$--C~ & ~C--C--H~ & ~C--C$'$--O~ & ~C$'$--O--H~ \\
 \hline
 ~~$\epsilon_{a}$~(eV)~~ &  1.3143  & 1.3        & 0.8     & 1.0     & 1.0 \\
 ~~$\varphi_0$~($^\circ$)~~& 120.0  & 109.5      & 120.0   & 109.5   & 108.5 \\
\hline
\end{tabular}
\end{center}
\end{table}

The interaction of two hydroxyl groups (hydrogen bond OH$\cdots$OH) was
described with the use of the potentials from the PCFF force field
\begin{equation}
W_{hb}=\sum_{i=1}^2\sum_{j=1}^2 \{\epsilon_{ij}[2(\bar{r}_{ij}/r_{ij})^9-3(\bar{r}_{ij}/r_{ij})^{12}]
+\kappa q_iq_j/r_{ij}\},
\label{f7}
\end{equation}
where $r_{ij}$ is the distance between the $i$-th atom of the first
and $j$-th atom of the second hydroxyl group, $q_i$ is the electric charge of $i$-th atom
(for oxygen atom $q_1 =-0.42e$, for hydrogen atom $q_2=0.42e$).
Distances $\bar{r}_{11}=3.58$, $\bar{r}_{12}=\bar{r}_{21}=3.19$, $\bar{r}_{22}=1.087$~\AA,
energy $\epsilon_{11}=0.0041629$, $\epsilon_{12}=\epsilon_{21}=0.0006938$,
$\epsilon_{22}=0.0003469$~eV, coefficient $\kappa=14.400611$~eV\AA/$e^2$.

\begin{figure*}[tb]
\begin{center}
\includegraphics[angle=0, width=0.8\linewidth]{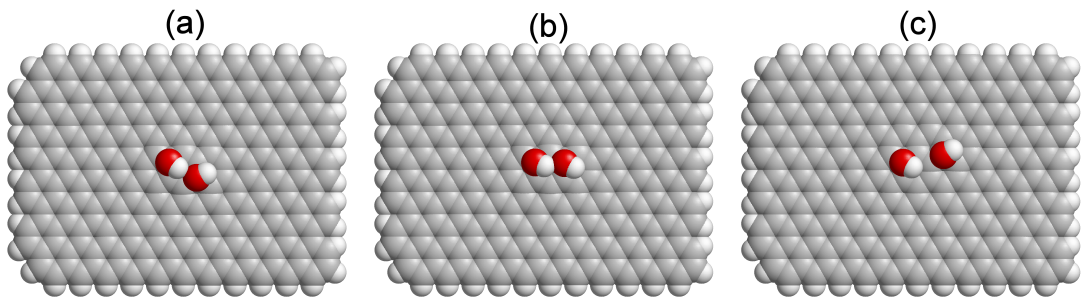}
\end{center}
\caption{\label{fig01}\protect
The  ground states of rectangular graphene sheet of size $2.83\times 1.99$~nm$^2$, placed on a flat
substrate (on the flat surface of graphite crystal) with two  valence-attached hydroxyl groups
(chemical formula is C$_{238}$H$_{42}$(OH)$_2$).
Part (a) shows the sheet when attaching hydroxyl groups to carbon atoms located at opposite
vertices of the same hexagon of the valence bonds of the sheet C--C,
(b) -- when attaching groups to atoms separated by two valence bonds,
(c) -- when attaching groups to atoms separated by four bonds.
Gray beads show carbon, white -- hydrogen, red -- oxygen atoms.
}
\end{figure*}

We define the interaction of the graphene sheet with a flat substrate using the potential $W_s(h)$,
which describes the dependence of the energy on the distance  $h$ of an atom to the substrate plane.
For a flat surface of a molecular crystal, the energy of the interaction of an atom
with a surface  can be described with a good accuracy by the ($k,l$)
Lennard-Jones potential \cite{skd2019prb}:
\begin{equation}
W_s(h)=\epsilon_s[k(h_0/h)^l-l(h_0/h)^k]/(l-k),
\label{f8}
\end{equation}
where $l>k$ is assumed for the exponents. The potential (\ref{f8}) has a minimum
of $W_s(h_0)=-\epsilon_s$ ($\epsilon_s$ is the binding energy of an atom with the substrate).
For a flat surface of crystalline graphite, the exponents in the potential (\ref{f8}) are
$l=10$, $k=3.75$. The binding energy is $\epsilon_s = 0.052$ and 0.0187~eV
for the C and  H atoms, respectively, and the corresponding equilibrium distances are
$h_0=3.27$ and 2.92~\AA.

\section{Chains of hydrogen bonds on one-side modified graphene sheet and nanotube}
\label{Sec2}

Consider a rectangular graphene sheet lying on a flat substrate.
To the edge atoms of the sheet, the hydrogen atoms are attached as shown in Fig.~\ref{fig01}.
To find the stationary state of one-side modified graphene sheet by hydroxyl groups,
it is necessary to find the minimum of the potential energy
\begin{equation}
E\rightarrow\min: \{ {\bf u}_n\}_{n=1}^N,
\label{f9}
\end{equation}
where $N$ is total number of atoms C, O, H on the sheet, ${\bf u}_n$ is a three-dimensional vector
defining position of $n$th atom, $E$ is a total potential energy of the molecular system
(given by the sum of all interaction potentials of atoms in the system (\ref{f2}),...,(\ref{f8})).
The minimization problem (\ref{f9}) is solved numerically by the conjugate gradient method.
Choosing the starting point of the minimization procedure, one can obtain all the main
stationary states of the modified sheet bonded with a flat substrate.

\begin{figure}[tb]
\begin{center}
\includegraphics[angle=0, width=0.8\linewidth]{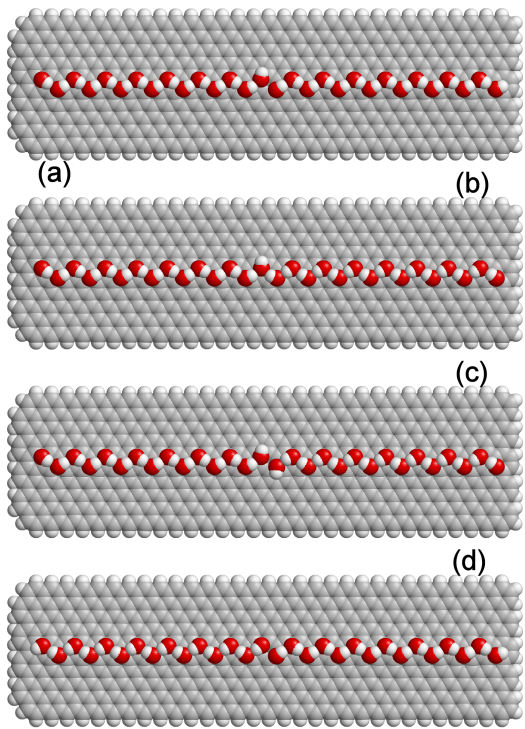}
\end{center}
\caption{\label{fig02}\protect
Stationary states of zigzag chain of hydrogen bonds formed by 30 hydroxyl groups OH
attached to the outer surface of graphene nanoribbon of size $7.78\times 1.99$~nm$^2$,
lying on a flat substrate (its chemical formula is C$_{638}$H$_{82}$(OH)$_{30}$):
(a) chain with one opened hydrogen bond;
chain with a positive orientation defect with (b) an integer and (c) a half-integer center
of symmetry; (d) chain with a negative orientation defect.
}
\end{figure}

\begin{figure}[tb]
\begin{center}
\includegraphics[angle=0, width=0.5\linewidth]{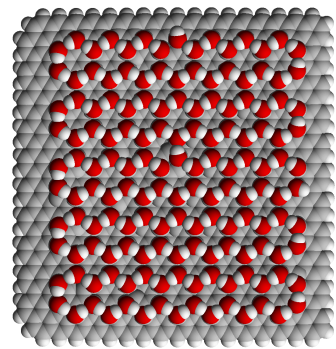}
\end{center}
\caption{\label{fig03}\protect
Stationary state of a square graphene sheet of size $4.39\times 4.62$~nm$^2$, lying on a flat
substrate with one-side attached 150 hydroxyl groups (C$_{790}$H$_{68}$(OH)$_{150}$).
Oxygen atoms are attached to carbon atoms forming a hexagonal lattice on the sheet
(neighboring atoms of the lattice are located at opposite vertices of the hexagons of the valence
bonds of the sheet).
At the center of the sheet there are stationary orientation defects (in the area of defects,
one oxygen atom participates in the formation of only one hydrogen bond, whereas
the neighboring atom takes part in the formation of three bonds).
Stationary states with open hydrogen bonds (defect energies are 0.161 and 0.174 eV)
are shown at the right and at the upper edge.
}
\end{figure}

Stationary states of a graphene sheet lying on a flat substrate with the hydroxyl groups OH
covalently attached to its outer side are presented in Fig. \ref{fig01}, \ref{fig02}, \ref{fig03}.
Solution of the problem (9) has shown that two hydroxyl groups form an energetically more
advantageous structure when they are attached to carbon atoms of the graphene sheet which
form opposite vertices of one hexagon of valence bonds C--C -- see Fig.~\ref{fig01}(a).
The distance between these atoms before attaching $r_{cc}=2.81$~\AA,
after attaching of the hydroxyl groups $r_{cc}'=2.90$~\AA,
the distance between oxygen atoms $r_{oo}=3.02$~\AA.
The attached hydroxyl groups form a hydrogen bond OH$\cdots$OH.
If the carbon atoms to which the hydroxyl groups are attached are separated
by two valence bonds C--C (see Fig.~\ref{fig01}(b), distance $r_{cc}=2.46$~\AA,
$r_{cc}'=2.59$\AA, $r_{oo}=2.84$~\AA)
or by four bonds (see Fig.~\ref{fig01}(c), distance $r_{cc}=4.26$~\AA,
$r_{cc}'=3.83$\AA, $r_{oo}=3.93$~\AA), in that case obtained structures will have higher energy.
Structure (b) is higher in energy then structure (a) in $\Delta E=0.0753$~eV,
and structure (c) -- in $\Delta E=0.1508$~eV (see Fig.~\ref{fig01}).
The analysis of stationary states shows, that the energy of the hydrogen bond between
two hydroxyl groups [see Fig.~\ref{fig01} (a)] $E_{hb}=0.12$~eV.
The value of this energy is small due to the fact that hydrogen atom in the bond O--H$\cdots$O
is not located on the straight line connecting oxygen atoms, deflection angle $\angle$HOO$=21.6^\circ$.

When hydroxyl groups attached to carbon atoms form the chain of opposite vertices
of hexagons of valence bonds of the sheet, the groups will form a zigzag chain
of hydrogen bonds (\ref{f1}).
Let us consider such chain located along a carbon nanoribbon lying on a flat substrate
-- see Fig.~\ref{fig02}. In this chain, the distance between neighboring oxygen atoms
(chain pitch) $r_{oo}=2.89$~\AA, the angle of the zigzag chain $\angle$OOO$=118^\circ$.
Such chains can have the following localized defects: breaking of one bond,
positive orientation defects with an integer and a half-integer center
and negative orientation defect -- see Fig.~\ref{fig02}.
The dissociation energy of one hydrogen bond
by the 120$^\circ$ rotation of one hydroxyl group $\Delta E_1=0.16065$~eV
(difference between the value of energy
of the stationary state of the chain with the defect and the value of energy
of the chain without a defect).

A positive orientation defect emerges when hydroxyl groups are directed toward each other
in the first and in the second part of the chain -- see Fig.~\ref{fig02} (b) and (c).
In the area of the defect localization the density of protons is excessive,
therefore it has an effective positive electric charge.
The defect can have two stationary states with integer and half-integer center of symmetry,
the energy of the first state $\Delta E_2=0.19152$~eV, the energy of the second $\Delta E_3=0.20154$~eV.

A negative orientation defect emerges when hydroxyl groups are directed oppositely
in the first and in the second part of chain -- see Fig.~\ref{fig02} (d).
In the area of the defect localization the density of protons is insufficient,
therefore it has an effective negative electric charge.
The energy of the stationary defect $\Delta E_4=0.28971$~eV.

On the graphene sheet, carbon atoms located on opposite vertices of hexagons of valence bonds
C--C build a hexagonal lattice with spacing $a=2r_0$, where $r_0$ is the length of the valence bond.
Each fourth atom of the sheet takes part in the formation of this lattice.
After one-side attaching of hydroxyl groups OH to these carbon atoms we will have a structure
with the chemical formula C$_4$(OH) (one hydroxyl group per four carbon atoms).
An example of such structure, namely a graphene sheet of size $4.39\times 4.62$~nm$^2$
(number of carbon atoms $N_c=790$, number of edge hydrogen atoms $N_h=68$,
number of hydroxyl groups $N_o=150$) is shown in Fig.~\ref{fig03}.

Solution of minimization problem (\ref{f9}) shows that in the case of a one-sided
chemical modification of graphene sheet lying on a flat substrate (namely on the flat surface
of a graphite crystal) the sheet maintains its flat form.
The oxygen atoms on the surface of the sheet form a hexagonal lattice that allows
hydroxyl groups to form chains of hydrogen bonds in different ways by rotating OH groups
around C--O bonds. In the ground state of this structure each oxygen atom must take part
in the formation of two hydrogen bonds (in one incoming and in one outcoming).
If neighboring oxygen atoms participate in the formation of hydrogen bonds,
than the distance between them is $r_{oo}=2.87$~\AA, and if they are not involved -- $r_{oo}=3.02$~\AA.
Let us note that in formed hydrogen bond O--H$\cdots$O the hydrogen atom is not located
strictly on the bond line (deflection angle $\angle$HOO$=17^\circ$).

The described system of hydrogen bonds allows the existence of two types of orientation defects.
In the localization region of positive defect, the oxygen atom takes part in the formation
of three hydrogen bonds; in the localization region of negative defect, the oxygen atom
participates in the formation of only one bond.
To form a pair of such defects, one hydroxyl group must be rotated by 120 degrees
around valence bond C--O  -- see Fig.~\ref{fig03}. After this pair of defects is formed,
the energy of the sheet increases by $\Delta E_d=0.123$~eV.
On the edges of the sheet, links with open hydrogen bonds may form;
the energy of such defects $\Delta E_o=0.161$, 0.174~eV.
\begin{figure}[tb]
\begin{center}
\includegraphics[angle=0, width=0.65\linewidth]{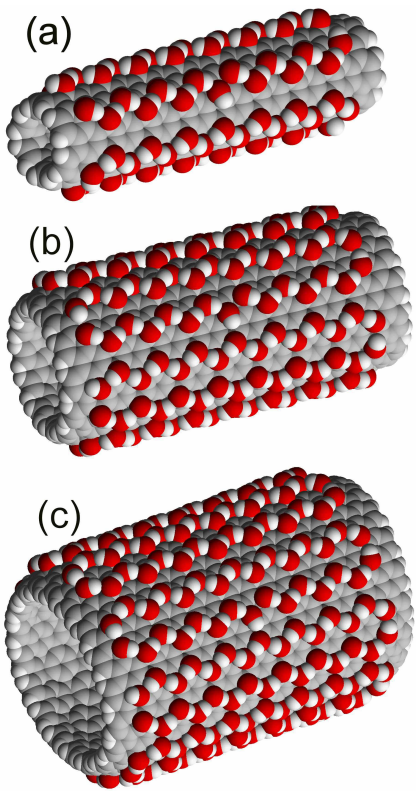}
\end{center}
\caption{\label{fig04}\protect
View of open armchair carbon nanotubes of length 4.13~nm with a chirality index:
(a)   (6,6), C$_{408}$H$_{24}$(OH)$_{90}$;
(b) (12,12), C$_{816}$H$_{48}$(OH)$_{180}$;
(c) (18,18), C$_{1224}$H$_{72}$(OH)$_{270}$.
Shown are the stationary states of the lattice of hydrogen bonds of hydroxyl groups
with an orientation defect in the middle of the nanotube and with one open edge bond
at the left edge.
}
\end{figure}

Similar molecular structures with chains of hydrogen bonds are obtained if hydroxyl groups
are covalently attached to the outer surface of the carbon nanotube to the carbon atoms
located at opposite vertices of hexagons of the C--C valence bonds.
Stationary states of modified nanotubes also can be found numerically as solutions
to the minimum potential energy problem (\ref{f9}) (in calculating of the total energy
the interactions with the substrate (\ref{f8}) may be omitted).

When the modification of the outer side has a homogenous character, the cylindrical shape
of the nanotube remains stable. The curvature of the surface of the nanotube leads
to a weakening of hydrogen bonds directed across the nanotube -- see Fig.~\ref{fig04}.
Therefore, the energy of defects of hydrogen bond chains depends on the diameter (chirality index)
of the nanotube.
Energies of the defects: $\Delta E_d=0.106$ and $\Delta E_o=0.094$ for nanotube with index (6,6);
0.110 and 0.115 for nanotube (12,12); 0.111 and 0.122~eV for nanotube (18,18).
\begin{figure}[tb]
\begin{center}
\includegraphics[angle=0, width=0.9\linewidth]{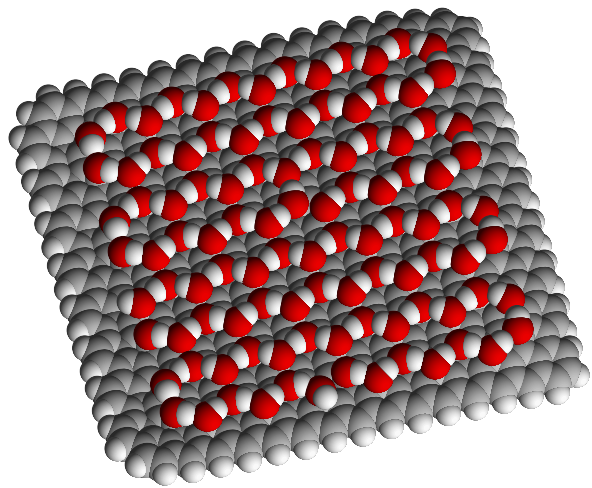}
\end{center}
\caption{\label{fig05}\protect
Stationary state of square graphene sheet of size $3.86\times 3.78$nm$^2$ with 208
hydroxyl groups OH attached to both sides of the sheet (C$_{574}$H$_{66}$(OH)$_{208}$).
Shown are the orientation defects of hydrogen bonds lattice in the center of the sheet
(in the area of defects, one oxygen atom participates in the formation of only one hydrogen bond,
whereas the neighboring atom takes part in the formation of three bonds).
The lower and left sides of the sheet show defects with open hydrogen bonds
(defect energies are 0.282 and 0.164~eV).
}
\end{figure}

\section{Chains of hydrogen bonds on two-side modified graphene sheet}
\label{Sec3}

By one-side modification, the graphene sheet maintains its stable flat shape only due to its
interaction with the flat substrate. Without a substrate, the modified sheet begins to bend,
and then it folds into roll shaped structures \cite{ssm2018prb}.
The sheet can retain its flat shape only by a symmetrical modification of its  both sides.
Stationary states of the sheet which was modified on both sides were found numerically as
solutions of the minimum potential energy problem (\ref{f9}) without
interaction with the substrate (\ref{f8}).
\begin{figure}[t]
\begin{center}
\includegraphics[angle=0, width=1.0\linewidth]{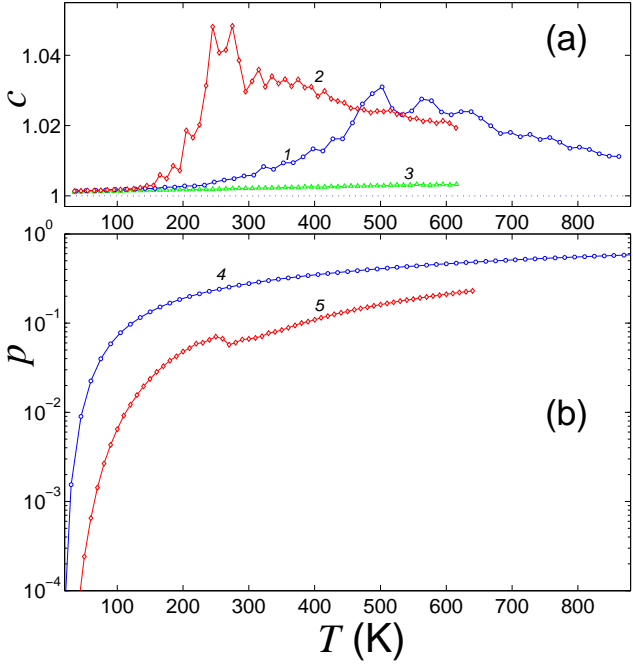}
\end{center}
\caption{\label{fig06}\protect
The dependence of (a) dimensionless heat capacity $c$ and (b) of the fraction of weakened hydrogen
bonds $p$ on the temperature $T$ for one-side modified graphene nanoribbon of size
$74.48\times 1.98$~nm$^2$ (C$_{6038}$H$_{622}$(OH)$_{300}$)
and square sheet of size $8.96\times 9.02$~nm$^2$ (C$_{3022}$H$_{154}$(OH)$_{660}$)
(curves 1, 4, and 2, 5). The nanoribbon and sheet are lying on a flat substrate.
Curve 3 shows the temperature dependence of the heat capacity for unmodified graphene sheet.
}
\end{figure}
\begin{figure}[t]
\begin{center}
\includegraphics[angle=0, width=1.0\linewidth]{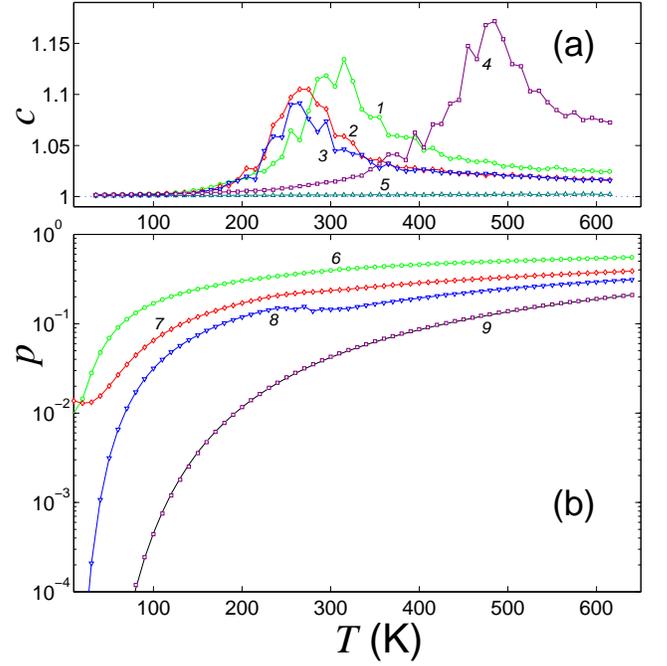}
\end{center}
\caption{\label{fig07}\protect
The dependence of (a) dimensionless heat capacity $c$ and (b) of the fraction of weakened hydrogen
bonds $p$ on the temperature $T$ for open single-wall armchair nanotube with modified outer
surface at: nanotube chirality index (6,6), length $L=25.7$~nm (C$_{2472}$H$_{24}$(OH)$_{600}$);
(12,12), $L=13.2$~nm (C$_{2544}$H$_{48}$(OH)$_{660}$);
(18,18), $L=13.2$~nm (C$_{3816}$H$_{72}$(OH)$_{918}$) (curves 1, 6; 2, 7; 3, 8).
Curves 4 and 9 show dependencies for graphene sheet modified on both sides
of the size $8.96\times 9.02$~nm$^2$ (C$_{3022}$H$_{154}$(OH)$_{1320}$).
The solid line connecting markers 9 shows the dependency $p(T)=2.095\exp(-212/T^{0.7})$.
Curve 5 shows the dependence $c(T)$ for the unmodified nanotube with index (6,6).
}
\end{figure}

When hydroxyl groups OH are two-side attached to carbon atoms located at opposite vertices
of hexagons of C--C valence bonds (these atoms form two hexagonal lattices, one for each side),
a hydroxygraphene structure with the chemical formula C$_2$(OH) emerges (one hydroxyl group
per two carbon atoms). An example of such structure, a graphene sheet of size $3.86\times 3.78$~nm$^2$
(chemical formula is C$_{574}$H$_{66}$(OH)$_{208}$) is shown in Fig.~\ref{fig05}. As we can see
from the figure, the modified sheet retains its flat shape (there are only a slight bendings
at the edges of the sheet).

Oxygen atoms form a hexagonal lattice on each side of the sheet, which allows hydroxyl groups OH
to form chains of hydrogen bonds in various ways due to their rotation around C--O bonds.
In the ground state of the sheet, each oxygen atom must take part in the formation of two hydrogen bonds.
If neighboring oxygen atoms take part in the formation of a hydrogen bond,
the distance between them $r_{oo}=2.78$~\AA, and if they are not involved  -- $r_{oo}=3.21$~\AA.
In the case of the formation of the hydrogen bond O--H$\cdots$O the hydrogen atom are also
not located strictly on the bond line (on the line connecting the oxygen atoms),
the deflection angle $\angle$HOO$=15^\circ$. It should be noted that here the values of
$r_{oo}$, $\angle$HOO are smaller than in the case of the one-side modified graphene.
This shows that hydrogen bonds are stronger if the sheet is modified on two sides.

The systems of hydrogen bonds arising on each side of the sheet also allow the formation
of orientation defects and open states of edge bonds -- see Fig.~\ref{fig05}.
Here, the energy of the formation of a pair of defects $\Delta E_d=0.191$~eV,
and the energies of the open states of the edge hydrogen bonds $\Delta E_o=0.282$, 0.164~eV.
\begin{figure*}[t]
\begin{center}
\includegraphics[angle=0, width=0.95\linewidth]{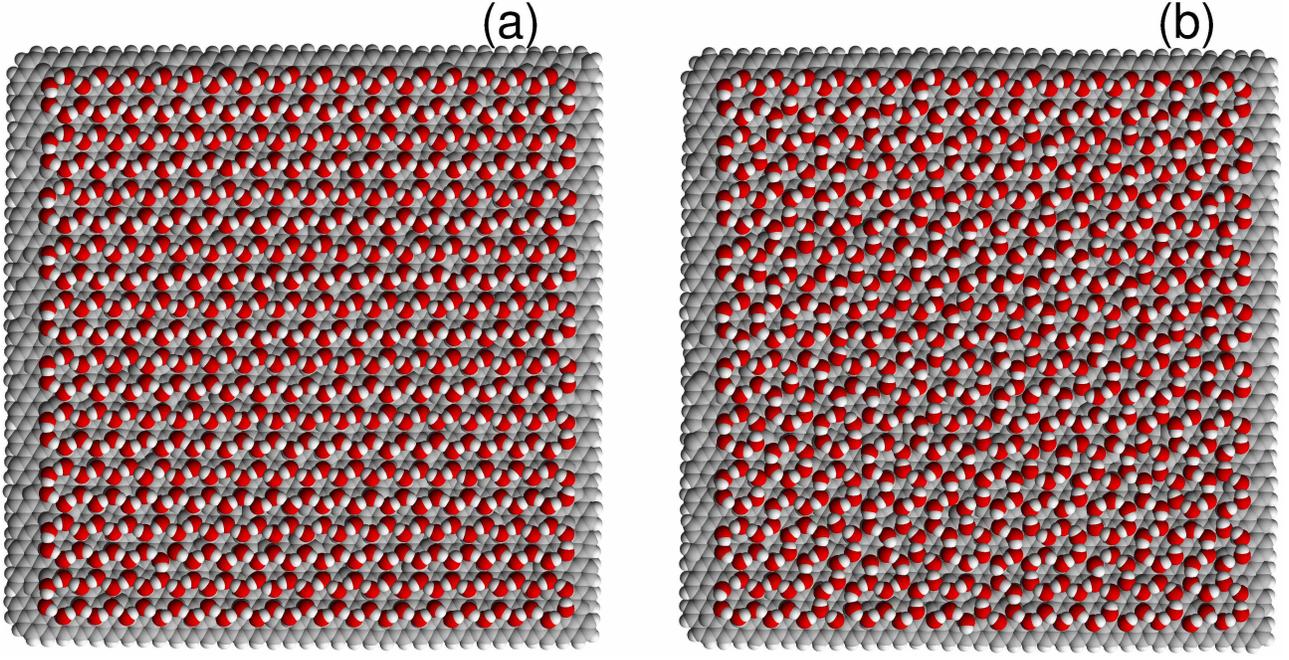}
\end{center}
\caption{\label{fig08}\protect
Configuration of one-side modified graphene sheet of size $8.96\times 9.02$~nm$^2$
(C$_{3022}$H$_{154}$(OH)$_{660}$) lying on a flat substrate
at temperature (a) $T=200$ and (b) $T=270$K.
}
\end{figure*}

\section{Dynamics of modified graphene sheets and nanotubes with HB networks}
\label{Sec3}

To simulate the dynamics of the thermalized sheet (nanotube),
the following system of Langevin equations was numerically integrated
\begin{equation}
M_n\ddot{\bf u}_n=-\frac{\partial~~~}{\partial{\bf u}_n}E-\Gamma M_n\dot{\bf u}_n-\Xi_n,~~n=1,...,N,
\label{f10}
\end{equation}
where $N$ is the total number of atoms in the molecular structure, $M_n$ is a mass of the $n$-th atom,
${\bf u}_n$ is a three-dimensional vector defining the coordinates of the $n$-th atom,
$\Gamma=1/t_r$ is the friction coefficient (the relaxation time  is $t_r=1$~ps),
$\Xi_n=\{\xi_{n,i}\}_{i=1}^3$ is a three dimensional vector of normally distributed random Langevin
forces with the following correlations:
$$
\langle\xi_{n,i}(t_1)\xi_{k,j}(t_2)\rangle=2M_nk_BT\Gamma\delta_{nk}\delta_{ij}\delta(t_1-t_2)
$$
($k_B$ is Boltzmann constant, $T$ is temperature of the Langevin thermostat).

As an initial condition for the equations of motion (\ref{f9}), we take the ground state
of the modified sheet (nanotube). Initially, the system of equations of motion was integrated over
the time $t_1=100t_r$. During this time, the molecular structure reaches its equilibrium with the
thermostat. Further integration allows us to analyze the dynamics of the thermalized macromolecules.
To this end, the temperature dependence of the average energy was found numerically,
$$
\langle H\rangle(T)=\lim_{t\rightarrow\infty}\frac{1}{t-t_1}\int_{t_1}^tH(\tau)d\tau,
$$
where the total energy of the molecular structure at time $t$ is
$$
H=\sum_{n=1}^N\frac12 M_n(\dot{\bf u}_n,\dot{\bf u}_n)+E.
$$
Then, the temperature dependence of the dimensionless heat capacity of the molecular structure was found
$$
c(T)=\frac{1}{3Nk_B}\frac{d~~}{dT}\langle H\rangle(T).
$$
The difference of the dimensionless heat capacity $c(T)$ from unity characterizes the nonlinearity
of the system dynamics.

Let us also define the fraction of weakened (broken) hydrogen bonds for each temperature
$$
p(T)=\langle (N_o-N_b)/N_o\rangle=\lim_{t\rightarrow\infty}\frac{1}{t-t_1}
\int_{t_1}^t(1-\frac{N_b(\tau)}{N_o})d\tau,
$$
where $N_b(t)$ is the number of hydrogen bonds,
$N_o$ -- the number of hydroxyl groups.
We assume that two hydroxyl groups form a hydrogen bond if their interaction energy
$W_{hb}<-0.08$~eV.
In the ground state of the sheet (at $T=0$K), the interaction energy of two hydroxyl groups
when they form the hydrogen bond $W_{hb}\approx -0.16$~eV,
and the interaction energy of other pair groups $W_{hb}>-0.08$~eV.
Therefore, in this case the number of bonds is equal to the number of hydroxyl groups
($N_b=N_o$) and the fraction of broken bonds $p=0$.
If the bonds are absent, i.e. when $N_b=0$, $p=1$.
In a general case, the fraction of broken bonds in HB networks may vary from zero to one
($0\le p\le 1$).
\begin{figure*}[t]
\begin{center}
\includegraphics[angle=0, width=0.95\linewidth]{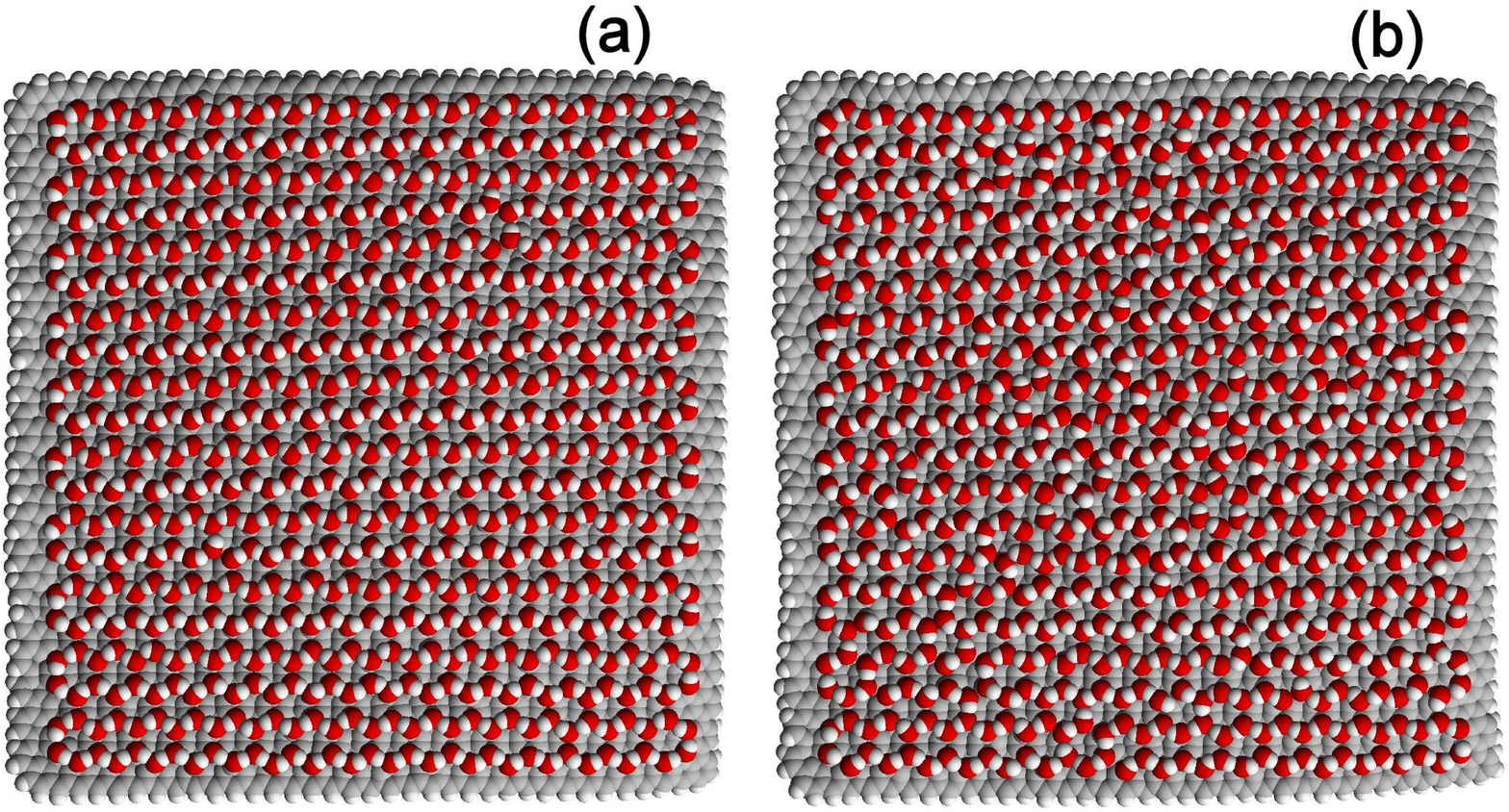}
\end{center}
\caption{\label{fig09}\protect
Configuration of two-side modified graphene sheet of size $8.96\times 9.02$~nm$^2$
(C$_{3022}$H$_{154}$(OH)$_{1320}$) at temperature (a) $T=300$ and (b) $T=500$K.
Only one sheet surface is shown.
}
\end{figure*}

To analyze the dynamics of a single chain of hydrogen bonds (\ref{f1}),
let us consider the graphene nanoribbon of size $74.48\times 1.99$~nm$^2$ lying on a flat substrate
with 300 hydroxyl groups attached to its outer side and forming a zigzag chain of hydrogen bonds
along its center (formula of the modified nanoribbon is C$_{6038}$H$_{622}$(OH)$_{300}$).
An example of such short nanoribbon is shown in Fig.~\ref{fig02}.
As the initial condition of a system of equations of motion (\ref{f10}) we take the ground state
of a nanoribbon without defects in the chain of hydrogen bonds.
For this purpose, we should first solve the minimum energy problem (\ref{f9}).
Then, we will numerically integrate the system of equations of motion (\ref{f10})
for different values of temperature of the Langevin thermostat $T$.

The dependencies of the dimensionless heat capacity of nanoribbon $c$
and the fraction of broken hydrogen bonds $p$ on temperature T are shown in Fig.~\ref{fig06}
(curves 1 and 4). As we can see, the fraction of broken bonds increases monotonically
with the increase of temperature. At low temperatures, $p$ grows exponentially fast,
at $T=300$K it reaches the value $p=0.27$, then the growth slows down (at $T=500$K $p=0.40$,
and at $T=800$K $p=0.55$). At $T<500$K, the dimensionless heat capacity of the nanoribbon $c$
increases monotonically with the increase of temperature, reaches its maximum value $c=1.031$
at $T=500$K, and then begins to decrease with the increase of temperature.
The increase in heat capacity is caused by the accumulation of orientational defects
in the hydrogen bond chain, whereas the decline is caused by the "melting"\ of the hydrogen bond
chain at high temperatures (at $T>500$K the orientational (torsional) motions of neighboring
hydroxyl groups become disconnected). In the absence of attached groups (i.e. in the absence
of chain of hydrogen bonds), the dimensionless heat capacity of the unmodified nanoribbon (sheet)
always remains near 1 (see Fig.~\ref{fig06}, curve 3).
\begin{figure*}[t]
\begin{center}
\includegraphics[angle=0, width=0.95\linewidth]{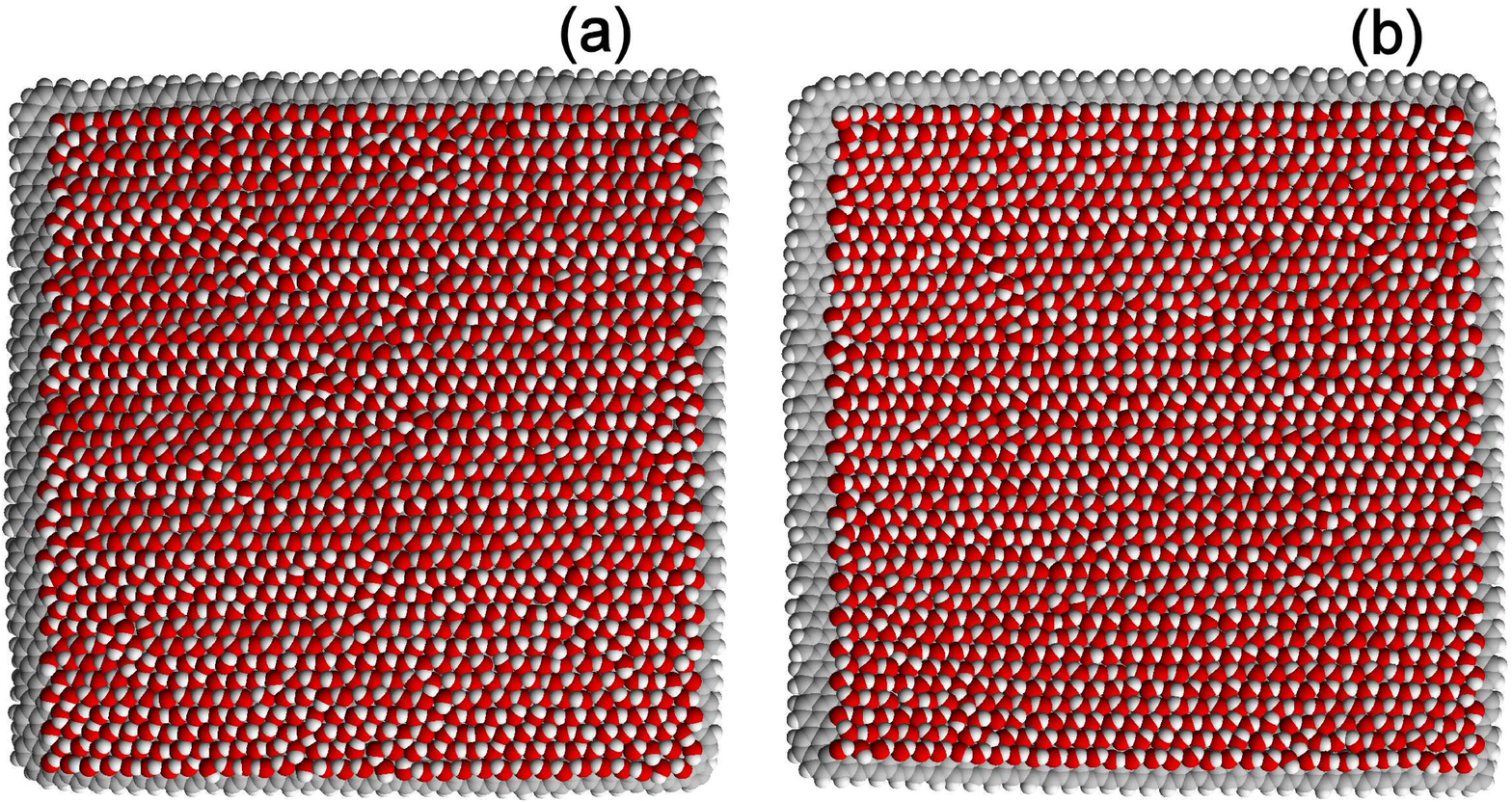}
\end{center}
\caption{\label{fig10}\protect
Configuration of two-side modified graphene sheet of size $ 8.96\times 9.02$~nm$^2$
with 2574 attached hydroxyl groups (C$_{ 3022}$H$_{154}$(OH)$_{ 2574}$,
the structure in the center of the sheet is described by the formula C(OH))
at temperature $T=600$K.
Part (a) shows the upper, part (b) -- the lower side of the sheet.
}
\end{figure*}

To analyze the dynamics of HB networks on the surface of a one-sided modified graphene sheet
with the general formula C$_4$(OH), let us consider the dynamics of a square sheet of size
$8.96\times 9.02$~nm$^2$ lying on a flat substrate with 660 hydroxyl groups covalently attached
to its outer side (the chemical formula of the sheet is C$_{3022}$H$_{154}$(OH)$_{660}$,
one OH group per four carbon atoms inside the sheet) -- see Fig.~\ref{fig08}.

The dependencies of the dimensionless heat capacity of the sheet $c$ and of the fraction
of weakened hydrogen bonds $p$ on temperature $T$ are shown in Fig.~\ref{fig06} (curves 2 and 5).
Numerical simulation of dynamics shows that at $T<250$K the heat capacity of the sheet increases
monotonically with increase of temperature, it reaches its maximum value $c=1.048$ at $T=275$K,
and then decreases.
An analysis of changes in the structure of HB networks shows that heat capacity increases
due to the accumulation of orientational defects in the system of hydrogen bonds.
For instance, at $T=200$K, the lattice of hydrogen bonds almost retains its ideal shape
(there is only one orientation defect in the lattice, namely in the lower left corner of the sheet),
and at $T=270$K, the hydrogen bond lattice already has a chaotic shape with many orientation
defects -- see Fig.~\ref{fig08}.

When $T<250$K, the fraction of weakened hydrogen bonds $p$ increases exponentially with
the increase of temperature, it reaches a local maximum $p=0.070$ at $T=250$K,
and at $T=270$K it reaches a local minimum $p=0.058$. Further increase in temperature
leads only to its slow growth (at $T=300$, 400, 600K $p=0.067$, 0.109, 0.211).

The simulation allows us to conclude that the increase in heat capacity of the one-sided
modified square graphene sheet C$_4$(OH) with the increase of temperature is caused
by the accumulation of orientation defects in the hydrogen bond lattice.
The maximum heat capacity is reached at $T=275$K, then the temperature increase is already
accompanied by the decrease in heat capacity, which is explained by the melting of the lattice
of hydrogen bonds (the movements of neighboring hydroxyl groups are becoming more and more disconnected).

To analyze the dynamics of HB networks on the modified outer surface of a carbon nanotube,
let us consider the dynamics of free open single-walled nanotubes, one with a chirality index
(6,6) of length $L=25.7$~nm (the formula of the modified nanotube is C$_{2472}$H$_{24}$(OH)$_{600}$),
another with index (12,12) of length $L=13.2$~nm (with the formula C$_{2544}$H$_{48}$(OH)$_{660}$)
and one of the same length with index (18,18) (C$_{3816}$H$_{72}$(OH)$_{918}$).
The structures of such modified nanotubes with length $L=4.13$~nm are shown in Fig.~\ref{fig04}.
As the initial condition of the system of equations of motion (\ref{f10}), we take the ground
state of a nanotube without defects in the lattice of hydrogen bonds.
For this purpose, we first solve the minimization problem (\ref{f9}).
Then we numerically integrate the system of equations of motion at different temperatures
of the Langevin thermostat.

The dependencies of the dimensionless heat capacity of nanotubes $c$ and the fraction
of weakened bonds $p$ on temperature $T$ are shown in Fig.~\ref{fig07} (curves 1, 2, 3 and 6, 7, 8).
As we can see, when temperature is low ($T<240$K), with the increase of temperature the fraction
of weakened hydrogen bonds grows exponentially quickly.
The heat capacity of the nanotube increases monotonically at low temperatures,
reaching its maximum value at temperature $T_0$, and then monotonically decreases.
For nanotube (6,6) temperature $T_0=315$K, the maximum value of the dimensionless heat capacity
$c_m=1.134$; for nanotube (12,12) temperature $T_0=270$K, $c_m=1.105$,
and for (18,18) $T_0=260$K, $c_m=1.092$.
Like for the sheets of modified graphene, the increase of the heat capacity of nanotubes
in this case is caused by the accumulation of orientation defects in the hydrogen bond lattice,
whereas the decrease of heat capacity at $T>T_0$ can be explained by the "melting"\
of the hydrogen bond lattice (at high temperatures, orientational motions of neighboring OH groups
are becoming more and more disconnected). For an ideal nanotube without attached groups,
the dimensionless heat capacity always remains near 1 (see Fig.~\ref{fig07}, curve 5).

To analyze the dynamics of HB networks on the surfaces of a two-side modified graphene sheet
with the general formula C$_2$(OH), let us consider the dynamics of free square sheet
of size $8.96\times 9.02$~nm$^2$ with 1320 OH groups attached to both sides of the sheet
(formula of the sheet is C$_{3022}$H$_{154}$(OH)$_{1320}$, there is one OH group per
two carbon atoms inside the sheet) - see Fig.~\ref{fig09}.

The temperature dependencies of the dimensionless heat capacity of sheet $c$ and the fraction
of weakened hydrogen bonds $p$ are shown in Fig.~\ref{fig07} (curves 4 and 9).
Numerical simulation of the dynamics shows that at $T<500$K the heat capacity of the sheet
increases monotonously with the increase of temperature, reaching a maximum value of
$c_m=1.13$ at $T=500$K and then decreasing.
Analysis of the structural changes in HB networks shows that here the increase of heat capacity
is also caused by the accumulation of orientational defects in the lattices of hydrogen bonds.
For instance, at $T=300$K the lattices of hydrogen bonds almost retains its ideal shape
(there are only two lattice defects on the upper right side of the sheet), whereas at
$T=500$K the hydrogen bond lattices already has a large number of orientation defects -- see Fig.~\ref{fig09}.

The fraction of weakened (open) hydrogen bonds $p$ increases exponentially with the increase
of temperature as function $p(T)=2.095\exp(-212 /T^{0.7})$ -- see Fig.~\ref{fig07} (curve 9).
At $T=100$, 300, 500K, the fraction of weakened bonds $p=0.0004$, 0.043, 0.138.

The simulation allows us to conclude that the two-side modification of the graphene sheet
with the formula C$_2$(OH) leads to the formation of a more rigid lattice of hydrogen bonds
in comparison to the one-sided modification C$_4$(OH). In the first case,
the "melting"\ of the lattice of hydrogen bonds begins at $T_0=500$K,
whereas in the second case it begins at $T_0=275$K.

As a comparison, let us also consider the modification of graphene with the most
dense attachment of hydroxyl groups, such as hydroxygraphane C$_2$H(OH) \cite{wk2010njp}.
In this structure hydrogen atoms H covalently attached to one side of the sheet like
in graphane sheet, whereas OH groups are attached to the other side.
Because of the asymmetric modification of the sides of the sheet, its flat shape is
not stable and the sheet necessarily folds into a roll structure.
Therefore, we will consider hydroxygraphene C(OH) with the most dense attachment of OH groups
to each side of the sheet (in this structure, there is one OH group per one carbon atom).

To analyze the dynamics of the modified sheet C(OH), we take a graphene sheet of size $8.96\times 9.02$~nm$^2$
with 2574 hydroxyl groups uniformly attached to each side of the sheet
(formula of the sheet is C$_{3022}$H$_{154}$(OH)$_{2574}$) -- see Fig.~\ref{fig10}.
As we can see from the figure, hydroxyl groups form a super-dense structure on each side of the sheet.
This structure almost does not allow the orientational mobility of OH groups.
Dynamics modeling has shown that at all temperatures $T<640$K the dimensionless heat capacity
of the modified sheet $c$ remains close to 1.
Thermal fluctuations practically do not lead to the formation of defects
in the super-dense lattices of hydroxyl groups.
Because of the low orientational mobility of OH groups, this structure is not well suited
for proton transport.

\section{Conclusions}

The numerical simulation of HB networks formed on the surface
of a graphene sheet during its functionalization by hydroxyl groups OH was carried out.
It was shown that two hydroxyl groups form the energetically most advantageous configuration when
they are covalently attached on one side of the sheet to carbon atoms forming opposite vertices
of the hexagon of graphene  C--C valence bonds (in this case, the attached groups form
hydrogen bond OH$\cdots$OH).
One-side attaching of OH groups to carbon atoms located at the opposite vertices of hexagons
of valence bonds leads to the emergence of hydroxygraphene C$_4$(OH).
The joined oxygen atoms on the outer side of the sheet lying on a flat substrate form a hexagonal
lattice and hydroxyl groups OH can in various ways form zigzag chains of hydrogen bonds
due to their rotation around C--O bonds. A two-sided modification of the sheet allows us to obtain
hydroxygraphene C$_2$(OH) with HB networks on each side of the graphene sheet.
We have also analyzed the possible stationary defects of such HB networks.

Modelling of dynamics of carbon nanotubes and graphene sheets modified by hydroxyl groups
has shown that their heat capacity increases monotonically with the increase of temperature
at low temperatures $T<T_0$, reaching a maximum value at $T=T_0$, and then monotonously decreases.
Analysis of changes in the structure of HB networks showed that the increase of heat capacity
is caused by the accumulation of orientation defects in the hydrogen bond lattice,
whereas the decrease of heat capacity at $T>T_0$ is explained by the "melting"\ of the lattice
(at high temperatures, orientational motions of neighboring groups become more and more independent).
For a single chain of hydroxyl groups attached to the outer side of the nanoribbon
the melting temperature $T_0=500$K, for one-side graphene modified sheet C$_4$(OH) temperature $T_0=260$K,
for carbon nanotubes modified from the outer side with a chirality index (6,6), (12,12), (18,18)
temperature $T_0=315$, 270, 260K.
The modification of both sides of the graphene sheet leads to the formation of more rigid lattices
of hydrogen bonds, for two-side modified sheet C$_2$(OH) the melting temperature of the lattices
of hydroxyl groups $T_0=485K$.

\section*{Acknowledgements}
The work was supported by the Russian Science Foundation
(award No. 16-13-10302). The research was carried out using
supercomputers at the Joint Supercomputer Center of the
Russian Academy of Sciences (JSCC RAS).

\section*{ORCIDiDs}
Alexander V. Savin --  https://orcid.org/0000-0003-0147-3515

\end{document}